\documentclass[a4, 14pt]{article}
\usepackage{amssymb,epic,eepic} %
\usepackage{amsfonts,amssymb,amsmath,latexsym}
\usepackage[latin1]{inputenc}
\author{ Nihat Ay $^{1,2}$, Markus M\"uller $^{1,3}$,
  Arleta Szko\l a $^{1}$\\ \\
$^{1}${\footnotesize Max Planck Institute for
Mathematics in the Sciences}\\
{\footnotesize Inselstrasse 22, 04103 Leipzig, Germany} \\
{\footnotesize e-mail:  nay@mis.mpg.de, szkola@mis.mpg.de}\\ \\
$^{2}${\footnotesize Santa Fe Institute} \\
{\footnotesize 1399 Hyde Park Road, Santa Fe, New Mexico 87501, USA}\\ \\
$^{3}${\footnotesize Institute
of Mathematics 7-2, Technical University Berlin} \\
{\footnotesize Stra\ss e des 17. Juni 136, 10623 Berlin, Germany }
}

\title{Effective complexity of stationary process realizations}
\date{April 4, 2011}

\newcommand{\s}{{\{0,1\}^{*}}}

\def\N{\mathbb {N}}
\def\Z{\mathbb {Z}}
\def\E{{\mathbb{E}}}
\def\EC{{\mathcal E}}
\def\Q{\mathbb {Q}}
\def\S{{\{0,1\}^{\infty}}}

\newtheorem{theorem}{Theorem}[section]         
\newtheorem{proposition}[theorem]{Proposition} 

\begin{document}

\maketitle

\begin{abstract}
 The concept of effective complexity of an object as the minimal
 description length of its regularities has been initiated by
 Gell-Mann and Lloyd. The regularities are modeled by means of
 ensembles, that is probability distributions on finite binary
 strings. In our previous paper \cite{ECieee} we propose a definition
 of effective complexity in precise terms of algorithmic information
 theory. Here we investigate the effective complexity of binary
 strings generated by stationary, in general not computable,
 processes. We show that under not too strong conditions long typical
 process realizations are effectively simple. Our results become most
 transparent in the context of coarse effective complexity which is a
 modification of the original notion of effective complexity that uses
 less parameters in its definition. A similar modification of the
 related concept of sophistication has been suggested by Antunes and
 Fortnow.
\end{abstract}


\section{Introduction}

The concept of effective complexity has been initiated by Gell-Mann and
Lloyd in \cite{EffectiveComplexity}, see also \cite{Gell-Mann96}. The
main motivation was to define a complexity measure that distinguishes
between regular and random aspects of a given object typically encoded
as a binary string. This is in contrast to Kolmogorov complexity which
is not sensitive to the source of incompressibility and in this sense
fails to capture what is meant by complexity in the common language.

The main idea underlying the concept has been considered at different
places in the literature, see \cite{rissanen},
\cite{GacsTrompVitanyi}, \cite{Vitanyi}, \cite{koppel},
\cite{vereshchagin}, \cite{soph}. It may be summarized as follows. One
considers programs computing a given binary string as consisting of
two parts: the implementation of an algorithm and a valid input for
that algorithm. Then the corresponding measures of complexity refer to
the algorithm part.

In \cite{EffectiveComplexity} the algorithm part has been motivated as
a description of a physical theory represented by a probability
distribution on finite binary strings while the second part has been
used to distinguish one among all possible objects contained in the
(typical) support of the distribution. Effective complexity is equal
to the length of the algorithm/theory part which is minimized over the
set of programs that compute the string and that are almost minimal,
i.e. their length is close to the Kolmogorov complexity of the string.

In \cite{ECieee} we have proposed a definition of effective complexity
in precise terms of algorithmic information theory. Our formalization
allows to include the concept into the context of algorithmic
statistics, which also deals with two-part codings of binary strings,
cf. \cite{GacsTrompVitanyi}. Instances of corresponding measures of
complexity are Kolmogorov minimal sufficient statistics and
sophistication, cf. \cite{GacsTrompVitanyi} and \cite{soph}. Roughly
speaking, while Kolmogorov minimal sufficient statistics of a binary
string $x$ is the minimal algorithmic statistic of $x$ from the model
class of finite sets, and sophistication refers to the model class of
total programs, effective complexity mainly coincides with the length
of algorithmic statistics of $x$ minimized over computable probability
distributions. For a more detailed presentation of the correspondence
relations between the different complexity measures we refer to our
previous paper \cite{ECieee}.

More precisely, the minimization domain of effective complexity
consists of computable probability distributions with total
information which is approximately equal to the Kolmogorov complexity
of the string: the tolerance level being specified by a parameter
$\Delta$. Total information has been defined by Gell-Mann and Lloyd in
\cite{Gell-Mann96} and \cite{EffectiveComplexity} as the sum of Kolmogorov
complexity and Shannon entropy of a given computable ensemble. It is
worth mentioning that it is equivalent to the concept of physical
entropy introduced by Zurek for large physical systems such as
thermodynamic engines, \cite{zurek}.

Restricting the minimization
domain of effective complexity by intersecting with subsets
corresponding to pre-knowledge about the object, which is subjective
to the observer, one ends up with a version of effective complexity
with constraints. As far as we know, there is no literature other than
the papers by Gell-Mann and Lloyd, cf. \cite{EffectiveComplexity},
where the idea to incorporate subjective pre-knowledge into the
measure of complexity has been considered explicitly.

Compared to the effective complexity without constraints, which we
will refer to as plain effective complexity or simply effective
complexity, this gives a larger value and is the reason why Gell-Mann
and Lloyd suggest to use the constrained version instead of the plain
one: ``If we impose no other conditions, every entity would come out
simple!'', see \cite{EffectiveComplexity}, page 392. This statement
has to be contrasted with the fact that there exist strings with large
plain effective complexity, cf. Theorem 13 of our previous work
\cite{ECieee}. See also corresponding results in the context of
algorithmic statistics and sophistication, Theorem 2.2 in
\cite{GacsTrompVitanyi} and Theorem 6.5 in \cite{Vitanyi},
respectively. Hence, the above conviction can be substantiated only in
a weaker version refering to some typical behaviour. In the present
contribution we find a framework to this end in the form of almost
sure statements in terms of probability theory.

In more detail, we investigate discrete-time stochastic processes with
binary state space in the context of effective complexity as it has
been defined in \cite{ECieee}. In addition to proving that typical
strings are simple with respect to the plain effective complexity, our
results also allow a deeper understanding of the dependence of
effective complexity on the parameter $\Delta$. Recall that this
parameter determines the minimization domain consisting of computable
ensembles with a total information that is $\Delta$-close to the
Kolmogorov complexity of the string. A corresponding parameter also
appears in the context of sophistication and, more generally,
algorithmic sufficient statistics, cf. \cite{soph},
\cite{GacsTrompVitanyi}. Conceptually, it also corresponds to the
significance level of Bennett's logical depth defined in
\cite{bennett}. The relation between effective complexity and logical
depth has been elaborated in
\cite{ECieee}.

In \cite{soph} Antunes and Fortnow suggested a modification of
sophistication called coarse sophistication. In an analogous way, we
introduce coarse effective complexity. It modifies the original
concept of plain effective complexity by, roughly speaking,
incorporating $\Delta$ into the definition as a further minimization
argument and as a consequence becoming independent of this
parameter. Our main results on effective complexity have direct
implications on the asymptotic behaviour of coarse effective
complexity. In particular, for an arbitrary stationary process the
value of coarse effective complexity of a typical finite string is
asymptotically upper bounded by any linear function of a string's
length.

After fixing notations and the mathematical framework in Section
\ref{Sec:Notations}, we formulate and prove our main result, Theorem
\ref{main-theorem}, in Section \ref{Sec:Stationary_processes}. It states that
sufficiently long typical strings generated by a stationary process
are effectively simple. The proof relies on the observation that the
total information of uniform distributions on universally typical
subsets is upper bounded by a value that exceeds the Kolmogorov
complexity of a typical string by any linear growing amount in the
string's length. In Section \ref{sec:coarse-ec} we introduce the
concept of coarse effective complexity. We show that strings of
moderate value of coarse effective complexity exist, see Theorem
\ref{thm:high-coarse-ec}, and derive from our main theorem an upper
bound on coarse effective complexity of long typical realizations of a
stationary process, see Theorem \ref{thm:low-coarse-ec}. Finally,
Section \ref{Sec:Conclusion} contains some conclusions and an outlook
for further analysis of effective complexity in its constrained
version.


\section{Notations and Preliminaries}\label{Sec:Notations}

We denote by $\s$ the set of finite binary strings, i.e. $\s=
\{ \lambda \} \cup \bigcup_{n \in \N} \{0,1\}^{n}$, where $\lambda$ is
the empty string, while the set of doubly infinite sequences $(\dots,
x_{-1}, x_0, x_1,\dots)$ with $x_i \in \{0,1\}$, $i \in \Z$, is
denoted by $\{0,1\}^{\infty}$. We write $\ell(x)$ for the length of
$x\in\s$.  Finite blocks $x_m^n=(x_m,x_{m+1},\dots, x_n)$, $m \leq n$,
of $x \in \S$ are elements of $\s$ of length $\ell(x_m^n)=n-m+1$. We
may identify them with cylinder sets $[x_m^n]:=\{y \in
\{0,1\}^{\infty}:\ y_i=x_i, m\leq i\leq n \}$. In a similar fashion
strings $x\in \s$ are associated to cylinder sets of the form
$[x]:=\{y \in \{0,1\}^{\infty}:\ y_i=x_i, 1\leq i\leq \ell(x) \}$. The
$\sigma$-algebra on $\{0,1\}^{\infty}$ generated by the cylinder sets
$[x_m^n]$, $m,n \in \Z$, $m\leq n$, is denoted by $\Sigma$. We write
$\mathcal{T}(\S)$ for the convex set of probability measures on $(\S,
\Sigma)$, which are invariant with respect to the left-shift $T$ on
$\S$. The subset of ergodic $T$-invariant probability measures, i.e.
the extremal points in $\mathcal{T}(\S)$, is denoted by
$\mathcal{E}(\S)$.

Let $P \in \mathcal{T}(\S)$. The random variables $X_i$, $i \in \Z$,
given by the coordinate projections $X_i(x):=x_i$, $x \in \S$,
respectively, represent a stationary process with values in $\{0,1\}$.
Typical outcomes of such stochastic processes are the main focus in
the present paper. The goal is to estimate their effective complexity.
We will refer to elements of $\mathcal{T}(\S)$ as stationary
probability measures and stationary stochastic processes
interchangeably.

Adopting the setup of \cite{ECieee} as far as possible we refer to
probability distributions on $\s$ as {\em ensembles}. 

For each $n \in \N$ we identify the joint distribution (alternatively
called $n$-block distribution) $P^{(n)}$ of $n$ successive outcomes
$(X_1,X_2, \dots, X_n)$ of a stationary process with an ensembles
$\E_n$ on $\s$ through the relation: $\E_n(x)=P^{(n)}(x)$ if
$\ell(x)=n$ and $\E_n(x)=0$ otherwise.

Recall the definition of {\em prefix Kolmogorov complexity} $K(x)$ of
a binary string $x \in \s$:
\begin{eqnarray*}
  K(x):= \min \{\ell(p):\ U(p)=x\},
\end{eqnarray*}
where $U$ is an arbitrary but fixed universal prefix computer. For
details concerning the basics as well as deeper results on Kolmogorov
complexity theory we refer to the book by Li and Vit\'anyi
\cite{LiVitanyi}.

We call an ensemble {\em computable} if there exists a program for the
universal computer $U$ that, given $x \in \s$ and $m \in \N$ as
inputs, computes an approximation of the probability $\E(x)$ with
accuracy of at least $2^{-m}$.

In \cite{ECieee} we have introduced an extension of the notion of
Kolmogorov complexity to the case of computable ensembles $\E$ with
computable and finite entropies $H(\E)$. Here we mean by entropy
$H(\E)$ the Shannon entropy of the probability distribution $\E$
defined by $-\sum_{x \in \s} \E(x)\log \E(x)$. Note that a computable
ensemble does not necessarily have a computable entropy, such that the
corresponding requirement is a restriction, see \cite{ECieee} for
details. In what follows a distinction only between general ensembles
and computable ones with computable and finite entropies is drawn. We
will refer to the latter ones as computable for short.

The {\em Kolmogorov complexity} $K(\E)$ of a computable ensemble $\E$
is defined as the length of the shortest computer program that, given
$x \in \s$ and $m \in \N$ as inputs, outputs both $\E(x)$ and $H(\E)$
with an accuracy of at least $2^{-m}$.

Additionally, we need to define computability of
stochastic processes. The following definition is a reformulation of
the notion of a ``computable measure'' in \cite{LiVitanyi}.

A stationary process $P$ is called {\em computable} if there exists a
program $p \in \s$ for a universal computer $U$ that, given $x \in
\{0,1\}^*$ and $m \in \N$ as inputs, computes the probability $P([x])$
up to accuracy $2^{-m}$.


\section{Effective Complexity of Stationary
  Processes}\label{Sec:Stationary_processes}

The goal is to show that under not too strong conditions long typical
samples of stationary processes are effectively simple. Before we make
rigorous statements we need a number of definitions. The first ones we
adopt from our previous paper \cite{ECieee}. 
\\ \\ 
Let $\delta \geq 0$. We say that an ensemble $\E$ is {\em
$\delta$-typical for a string} $x \in \s$, or alternatively, we call
$x$ $\delta$-typical for $\E$, if the Shannon entropy $H(\E)$ of $\E$
is finite and
\begin{eqnarray*} 
- \log \E(x) \leq H(\E)(1+\delta). 
\end{eqnarray*}

In particular, the special case of an equidistributed ensemble is
$\delta$-typical for all strings in the support and any $\delta \geq
0$.
\\ \\
The {\em total information $\Sigma(\E)$} of a computable ensemble $\E$
is defined by
\[
\Sigma(\E):= H(\E) + K(\E).
\]
For a motivation of the two above defintions see \cite{ECieee}.
\\ \\
Let $\delta \geq 0$ and $\Delta >0$. {\em Effective complexity }
$\EC_{\delta, \Delta}(x)$ of a finite string $x\in\s$ is defined by
\begin{eqnarray*} \EC_{\delta, \Delta} (x):= \min \{ K(\E):\ \E \in
\mathcal{P}_{\delta, \Delta}(x) \}, \end{eqnarray*} where
$\mathcal{P}_{\delta, \Delta}(x)$ denotes the {\em minimization
domain} associated to $x$: \begin{eqnarray}\label{def:min_domain}
\mathcal{P}_{\delta, \Delta}(x):=\{\E:\ \E\ \mbox{computable
ensemble},\E\ \delta\mbox{-typical for } x,\ \Sigma(\E) \leq K(x)+
\Delta\}.
\end{eqnarray}
We refer to elements of $\mathcal{P}_{\delta, \Delta}(x)$ as {\em
  effective ensembles} for $x$.
\\ \\
Using the viewpoint of \cite{EffectiveComplexity}, which was reviewed
in \cite{ECieee}, effective ensembles represent theories or
explanations that are judged to be good for $x$.
\\ \\
The more general notion of
{\em effective complexity with constraints} has been suggested in
\cite{EffectiveComplexity} mainly to circumvent problems of plain
effective complexity. We have discussed them shortly in the
Introduction. The main idea is that the constraints reflect some
pre-knowledge about the possible theory for $x$. In \cite{ECieee} we
have proposed a formalization of the constrained version in the
following manner:
\[
\EC_{\delta, \Delta}(x |\ \mathcal{C}):=\min \{ K(\E):\ \E \in
\mathcal{P}_{\delta, \Delta}(x), \E \in \mathcal{C} \},
\] 
where $\mathcal{C}$ is a subset of $\mathcal{P}(\s)$. Note that with
$\mathcal{C}=\mathcal{P}(\s)$ we have
$\EC_{\delta,\Delta}(x)=\EC_{\delta, \Delta}(x | \mathcal{C})$, for
all $x \in \s$.
\\ \\
In what follows other essential concepts are that of typical and/or
universally typical subsets.
\\ \\
Let $P$ be a $T$-invariant probability measure on
  $(\{0,1\}^{\infty}, \Sigma)$. We call a sequence of subsets $M_n\in
  \Sigma$, $n \in \N$, {\em $P$-typical} if
\[
    \lim_{n \to \infty} P(M_n)=1.
\]
We call $(M_n)$ {\em strongly $P$-
typical} if for $P$-almost all $x$
there exists an $N_x \in \N$ such that
\[
    x \in M_n\qquad\mbox{for every } n \geq N_x.
\]

The above notions of typicality apply
naturally to sequences $M_n \subseteq \{0,1\}^n$, $n \in \N$, if we
identify subsets $M_n$, $n \in \N$, with cylinder sets $ [M_n]
\subseteq \{0,1\}^{\infty}$, respectively.
\\ \\
Let $\Lambda$ be a set of stationary processes with values in
  $\{0,1\}$. We call $M_n \subseteq\{0,1\}^n$, $n \in \N$,
  \textit{universally typical for} $\Lambda$ if the sequence is
  $P$-typical for every $P \in \Lambda$, i.e. $ \lim_{n \to \infty}
  P^{(n)}(M_n)= 1$. We call the sequence \textit{universally strongly
    typical for} $\Lambda$ if it is strongly $P$-typical for every $P
  \in \Lambda$.
\\ \\
For sets $\Lambda_r \subseteq \mathcal{E}(\S)$ consisting of ergodic
processes with entropy rate upper bounded by $r > 0$ there exist
universally typical subsets $T_{r,n} \subseteq \{0,1\}^n$ with
\begin{eqnarray}\label{size-of-universal-subsets} 
|T_{r,n}| \leq 2^{rn}
\end{eqnarray}
for all $n \in \N$. Moreover, there are methods to construct such
sequences of universally typical subsets for $\Lambda_r$. We will
apply the Lempel-Ziv algorithm in the construction procedure below,
see \cite{kieffer, ziv, CoverThomas}. The main point is that all we
need to know about an ergodic process $P$ is its entropy rate $h_P$.
This allows to prove the following theorem for stationary, in
general {\it not computable} processes.
\begin{theorem}\label{main-theorem}
Let $P$ be a stationary
  process, $\delta \geq 0$, $\epsilon >0$ and $\Delta_n=\epsilon
  n$. Then $P$ is effectively simple in the sense that for $P$-almost
  every $x$,
\begin{eqnarray}\label{main-ineq}
	\EC_{\delta, \Delta_n}(x_1^n) \stackrel+ < \log n +
	\mathcal{O}(\log\log n).  
\end{eqnarray}
\end{theorem} 
{\it Proof.}  1. Assume that $P$ is an ergodic process with entropy
rate $h_P$. We construct universally typical subsets $T_{r,n}
\subseteq \{0,1\}^n$, $n \in \N$, such that for an appropriate choice
of the parameter $r= r(h_P,\epsilon)$ the total information of the
uniform distributions $\E_{r,n}$ on $T_{r,n} \subseteq \{0,1\}^n$,
respectively, is upper bounded by $K(x_1^n) + \Delta_n$ of $P$-almost
every $x \in \{0,1\}^{\infty}$ and for $n$ large enough. Hence the
Kolmogorov complexity of $\E_{r,n}$, which is approximately upper
bounded by $\log n$, gives an estimate from above on the effective
complexity $\EC_{\delta, \Delta_n}(x_1^n)$ of sufficiently long
$P$-typical strings $x_1^n$.

 First, let $r>0$ be arbitrary and define for each $n \in \N$ the
 subset $T_{r,n} \subseteq \{0,1\}^n$ as the set consisting of all
 binary strings $x_1^n$ which are mapped by the Lempel-Ziv (LZ)
 algorithm to a code word of length $\ell_{LZ}(x_1^n)$ lower than
 $nr$.  Then the sequence $T_{r,n}$, $n \in \N$, is universally
 typical for the set $\Lambda_r$ of ergodic processes with entropy
 rates lower than $r$.

 Recall the following remarkable property of the LZ algorithm: For
 every ergodic process $Q$ with entropy rate $h_Q$ it holds $ \lim_{n
   \to \infty}\frac{1}{n}\ell_{LZ}(x_1^n)= h_Q$ for $Q$-almost all $x
 \in \{0,1\}^{\infty}$. This implies that indeed the subsets $T_{r,n}$
 as constructed above are typical for any ergodic $Q$ with $h_Q <r$.

 The upper bound (\ref{size-of-universal-subsets}) on the size
 $|T_{r,n}|$ follows from the fact that the LZ algorithm is a faithful
 coder and as such satisfies the Kraft inequality:
\begin{eqnarray*}
  1&\geq& \sum_{x_1^n \in \{0,1\}^n} 2^{-\ell_{LZ}(x_1^n)} \geq
  \sum_{x_1^n \in T_{r,n}} 2^{-\ell_{LZ}(x_1^n)}\\
  &\geq& \sum_{x_1^n \in
    T_{r,n}} 2^{-nr}= |T_{r,n}|2^{-nr}.
\end{eqnarray*}

Next, we show that if $r$ is chosen to be a positive rational number
satisfying $0< r-h_P < \epsilon/4$ then for $P$-almost every $x \in
\{0,1\}^{\infty}$ there exists an $N_x \in \N$ such that
\begin{eqnarray}\label{bound-on-total-info-of-unif-distr}
\Sigma(\E_{r,n})\leq K(x_1^n)+
\Delta_n\qquad\mbox{for every } n \geq N_x,
\end{eqnarray}
where again $\E_{r,n}$ denotes the uniform distribution on the
universally typical subset $T_{r,n}$.  First note that for all $n
\in\N$
\begin{eqnarray*}
 H(\E_{r,n})= \log |T_{r,n}|\leq rn.
\end{eqnarray*}  
Secondly, we prove that there is a constant $c \in \N$ such that for
all $n \in \N$
\begin{eqnarray}\label{upper-bound-on-compl-of-unif-distr}
  K(\E_{r,n}) \leq K(n)+ K(r) + c.
\end{eqnarray}
This is derived from the existence of a program $p$ of length $c$
which expects as inputs $n \in \N$, $r \in \Q$ and $x \in \{0,1\}^*$
and outputs the value $\frac{1}{|T_{r,n}|} $ if $x \in T_{r,n}
\subseteq \{0,1\}^n$ and $0$ otherwise. Thus for fixed inputs $n$ and
$r$ it gives a description of the uniform distribution $\E_{r,n}$ on
$T_{r,n}$.

Indeed, $p$ may be constructed on the base of a program $p_{LZ}$
implementing the Lempel-Ziv (LZ) algorithm on the given reference
universal computer $U$. For given inputs $n$ and $r$ let $p$ apply
$p_{LZ}$ as a subroutine in order to determine elements of $T_{r,n}$.
Then for fixed $n \in \N$ the number $|T_{r,n}|$ and hence the
probability value $1/|T_{r,n}|$ of each $x \in T_{r,n}$ may be
calculated easily.

To specify $r \in \Q$ and $n \in \N$ a number of $K(r)+K(n)$ bits is
sufficient. With $c=\ell(p)$ the estimate
(\ref{upper-bound-on-compl-of-unif-distr}) follows.

Next, fix an $N \in \N$ such that $K(n) +K(r)+c \leq
\frac{\epsilon}{4} n$ for all $n \geq N$. 
Then
\begin{eqnarray}\label{Sigma-upper-bound2}
	\Sigma(\E_{r,n})\leq nr + K(n) +K(r)+c \leq
	n(r+\frac{\epsilon}{4})\leq n(h+\frac{\epsilon}{2}),
\end{eqnarray}
where the last inequality holds by assumption $r-h
<\frac{\epsilon}{4}$. According to the theorem by Brudno, see
\cite{Brudno}, for $P$-almost all $x$ there exists an $N_{x,\epsilon}
\in \N$ such that $K(x_1^n) \geq n(h -
\frac{\epsilon}{2})$ for all $n \geq N_{x,\epsilon}$. It follows for
$\Delta_n=\epsilon n$
\begin{eqnarray}\label{complexity-lower-bound2}
	K(x_1^n) + \Delta_n\geq n (h + \frac{\epsilon}{2}),\qquad n
	\geq N_{x,\epsilon}.
\end{eqnarray}
Relations (\ref{Sigma-upper-bound2}) and
(\ref{complexity-lower-bound2}) together imply
(\ref{bound-on-total-info-of-unif-distr}) for $P$-almost all $x$ and
$n \geq N_x:=\max \{ N_{x,\epsilon}, N \}$. It follows that $P$-almost
surely the effective complexity $\EC_{\delta,\Delta_n}(x_1^n)$ is
upper bounded by the Kolmogorov complexity of $\E_{r,n}$ for $n \geq
N_x$:
\begin{eqnarray*}
	\EC_{\delta,\Delta_n}(x_1^n) &\leq& K(\E_{r,n}) \leq
	K(n)+K(r)+c\\ &\stackrel + <& \log n + \mathcal{O}(\log\log
	n).
\end{eqnarray*}

2. Now let $P$ be an arbitrary stationary process. Recall that there
is a unique ergodic decomposition of $P$
\begin{eqnarray*}
  P=\int_{\mathcal{E}(\S)} Q d\mu(Q).
\end{eqnarray*} 
Moreover, to $P$-almost every $x\in \S$ we may associate an ergodic
component $Q_x$ of $P$ such that $x$ is a typical element of $Q_x$.
Then there exists an $N_{x,\epsilon}$ such that
\begin{eqnarray*}
	K(x_1^n)+\Delta_n \geq n(h_x + \frac{\epsilon}{2}),\qquad n
	\geq N_{x,\epsilon},
\end{eqnarray*}
where $h_x$ denotes the entropy rate of $Q_x$. Hence the proof for the
stationary case reduces to the ergodic situation considered in the
first part above. $\Box$


\section{Coarse Effective Complexity}\label{sec:coarse-ec}
Our main result becomes most transparent if presented in the context
of \textit{coarse effective complexity}. This is a modification of
plain effective complexity which incorporates the parameter $\Delta$
as a penalty into the original formula. It is inspired by a
corresponding modification of sophistication, called coarse
sophistication, which has been introduced by Antunes and Fortnow in
\cite{soph}.
\\ \\
Let $\delta \geq 0$. The \textit{coarse effective complexity}
$\EC_{\delta}(x)$ of a finite string $x \in \{0,1\}^*$ is defined by
\begin{eqnarray*}
	\EC_{\delta} (x):=\min \{K(\E)+ \Sigma(\E) - K(x): \E
	\mbox{ is computable ensemble, }\E\ \delta-\textrm{typical for }
	x\}.
\end{eqnarray*}

The term $\Sigma(\E) - K(x)$ accounts for the exact value by which the
total information of an ensemble $\E$ exeeds the Kolmogorov complexity
of $x$. By definition of total information $\Sigma(\E)$ an equivalent
expression for $\EC_{\delta}(x)$ reads
\begin{eqnarray*}
	\EC_{\delta}(x)= \min \{2K(\E)+ H(\E) - K(x): \E
	\mbox{ is computable ensemble, }\E\ \delta-\textrm{typical for }
	x\}.
\end{eqnarray*}
We derive the basic properties of coarse effective complexity
similarily as it has been done in \cite{soph} in the context of coarse
sophistication. That is firstly, in the proposition below, we prove an
upper bound on coarse effective complexity. Secondly, we show
existence of strings, which are close to saturate this bound.
\begin{proposition} Let $\delta \geq 0$. There is a constant $c$ such
that for all $x \in \{0,1\}^*$ we have
\begin{eqnarray}
	\EC_{\delta}(x)\leq \frac{n}{2} + \log n +c,
\end{eqnarray}
where $n = \ell(x)$.
\end{proposition}
{\it Proof.} Suppose that $K(x) \leq \frac{n}{2}+ \log n$. Let $\E_x$
denote the ensemble with $\E(x)=1$ and $\E(y)=0$ for $y \not= x$. Note
that $\E_x$ is trivially $\delta$-typical for $x$ for any $\delta \geq
0$ and obviously $H(\E_x)=0$. Moreover, there is a constant $c_1$ such
that it holds $K(\E_x) \leq K(x)+c_1$. This implies the upper bound
\begin{eqnarray}
	\EC_{\delta}(x) &\leq& 2 K(\E_x) +0 - K(x) \nonumber \\ &\leq&
	K(x) + 2 c_1\nonumber\\ &\leq& \frac{n}{2} + \log n + 2c_1,
	\nonumber
\end{eqnarray}
where the last line holds by assumption.

Now, suppose that $K(x) > \frac{n}{2} + \log n$. Let $\E_n$ be the
ensemble on $\{0,1\}^*$ given by $\E_n(y)= \frac{1}{2^n}$ for all $y$
with $\ell(y)=n$ and vanishing elsewhere. Then $H(\E_n)= n$ and there
exists a constant $c_2$, independent of $n$, such that $K(\E_n) \leq
\log n + c_2$. It follows
\begin{eqnarray*}
	\EC_{\delta}(x)&\leq& 2 \log n + 2 c_2 + n -K(x)\\ &\leq&
	\frac{n}{2} + \log n +2 c_2,
\end{eqnarray*}
where, again, the second line holds by assumption on $K(x)$.  Setting
$c:= \max\{2c_1, 2c_2\}$ completes the proof. $\Box$
\begin{theorem}\label{thm:high-coarse-ec}
Let $\delta \geq 0$. For every sufficiently large $n
\in \N$ there exists a string $x\in \{0,1\}^n$ with
\begin{eqnarray}
	\EC_{\delta}(x)\geq (1-3\delta) \frac{n}{2} - (2+3\delta) \log
	n -2\log \log n + C,
\end{eqnarray}
where $C$ is a global constant.
\end{theorem}
{\it Proof.} For $x \in \{0,1\}^*$ and $\Delta \geq 0$ denote by
$\E_{x}^{\Delta}$ the minimal ensemble associated to
$\EC_{\delta,\Delta}(x)$. Due to Lemma 22 in \cite{ECieee} for every
$\epsilon>0$ there exists a subset $S_x^{\Delta}$ of $\{0,1\}^*$ such
that
\begin{eqnarray}
	\log |S_x^{\Delta}| &\leq& H(\E_x^{\Delta})(1+ \delta) +
	\epsilon \label{upper-bound-on-size}\\ K(S_x^{\Delta})&\leq&
	K(\E_x^{\Delta}) + c_1
	\label{upper-bound-on-compl},
\end{eqnarray} 
where $c_1$ is a global constant. In \cite{ECieee} we have proven the
relation
\begin{eqnarray}\label{eqn31-in-ECieee}
	K(x| S_x^{\Delta}, K(S_x^{\Delta}))\geq \frac{\log
	|S_x^{\Delta}|}{1+ \delta} -\log n -2 \log \log n
	-\Lambda_{\Delta},
\end{eqnarray}
which holds for arbitrary $x \in \{0,1\}^n$, $n \in \N$. The term
$\Lambda_{\Delta}$ is constant in $x\in \{0,1\}^*$ and monotonically
increasing in $\Delta$, cf. $(32)$ in \cite{ECieee}. Now, let
$K_n:=\max \{K(t)|\ t \in \{0,1\}^n\}$ and define
\begin{eqnarray*}
	k:= n - \delta(K_n + \Delta_n + \epsilon) + \log n + 2 \log
	\log n - \Lambda_{\Delta_n}- c_2,
\end{eqnarray*}
where $\Delta_n:= \frac{n}{2}+ \log n + c$ is the upper bound on
$\EC_{\delta}(x)$ obtained in the previous proposition and $c_2$ is a
global constant from Theorem IV.2 in \cite{GacsTrompVitanyi}, see also
Lemma 12 in \cite{ECieee}. If $n$ is large enough then $0< k<n$ holds,
and Theorem IV.2 in
\cite{GacsTrompVitanyi} applies: There is a string $x_k \in \{0,1\}^n$ such that
\begin{eqnarray}
	K(x_k| S, K(S)) < \log |S| -n -k + c_2,
\end{eqnarray}
for every set $S \ni x_k$ with $K(S) < k-c_3$, where $c_3$ is another
global constant. Let $\E_{x}$ denote the minimizing ensemble
associated to coarse effective complexity $\EC_{\delta}(x)$ and
$\Delta_x:=K(\E_x) +H(\E_x) -K(x)$ such that $\E_{\delta}(x)= K(\E_x)
+ \Delta_x$. Further, define $S_x:= S_x^{\Delta_x}$. It holds the
inequality
\begin{eqnarray*}
	-\delta(K_n + \Delta_n + \epsilon) &\leq& -\delta (K(x_k) +
	\Delta_x +\epsilon)\\ &\leq&
	-\delta\left(H(\E_{x_k})+\epsilon \right)
	\nonumber\\ &\leq& -\delta\left( H(\E_{x_k})+
	\frac{\epsilon}{1+\delta}\right)\nonumber \\ &=&
	\frac{-\delta}{1+\delta}
	\left(H(\E_{x_k})(1+\delta) + \epsilon
	\right)\nonumber \\ &\leq& \left( \frac{1}{1+\delta} -1
	\right) \log |S_{x_k}|,
\end{eqnarray*}
where the last upper bound holds by (\ref{upper-bound-on-size}). Now
suppose that $K(S_{x_k}) < k -c_1$. Then
\begin{eqnarray*}
	K(x_k| S_{x_k}, K(S_{x_k})) &<& \log |S_{x_k}|-n+k+c_2\\
	&\leq& \log |S_{x_k}|-\log n - 2 \log \log n \\ &&
	-\Lambda_{\Delta_n}- \delta(K_n + \Delta_n +\epsilon)\\ &\leq&
	\frac{\log |S_{x_k}|}{1+\delta} -\log n - 2\log \log n\\ &&
	-\Lambda_{\Delta_n}\\ &\leq& \frac{\log |S_{x_k}|}{1+\delta}
	-\log n - 2\log \log n\\ && -\Lambda_{\Delta_{x_k}}.
\end{eqnarray*}
But the strict inequality is a contradiction to
(\ref{eqn31-in-ECieee}). Hence our assumption must be false and we
instead have $K(S_{x_k}) \geq k-c_3$. By $\EC_{\delta}(x) =
K(\E_x)+\Delta_{x}$ and using both (\ref{upper-bound-on-compl}) and
the bound $K_n \leq n + 2\log n + \gamma$, where $\gamma$ is a global
constant, we finally obtain
\begin{eqnarray*}
	\EC_{\delta}(x_k)&=& K(\E_{x_k})+ \Delta_{x_k}\\ &\geq&
	K(S_{x_k})-c_1+ \Delta_{x_k} \\ &\geq& k-c_3-c_1 +
	\Delta_{x_k}\\ &\geq& n -\delta(\frac{3}{2}n+3 \log n + \gamma
	+c+ \epsilon)-\log n\\ & & - 2\log \log n - \frac{n}{2} - \log
	n - c- c_2 -1 -c_3 -c_1\\ &=& (1-3\delta)\frac{n}{2}-(2
	+3\delta)\log n -2\log \log n + C,
\end{eqnarray*}
where $C:= -\delta(\gamma + \epsilon)-1- (1+\delta)c-c_1-c_2-c_3$.
$\Box$
\\ \\
Although, according to the above theorem, for arbitrary large $n$ the
existence of strings of length $n$ with moderate coarse effective
complexity is ensured, the coarse effective complexity of sufficiently
long prefixes of a typical stationary process realization becomes
small. This is a direct implication of Theorem \ref{main-theorem}. 
\begin{theorem}\label{thm:low-coarse-ec}
Let $P$ be a stationary process, $\delta \geq 0$ and
$\epsilon > 0$. Then for $P$-almost every $x$
\begin{eqnarray}\label{implication-of-Thm10}
	\EC_{\delta}(x_1^n)\leq \epsilon n + \log n +
	\mathcal{O}(\log \log n).
\end{eqnarray}
\end{theorem}
{\it Proof.} By definiton of coarse effective complexity it holds
$\EC_{\delta}(x) \leq \Delta + \EC_{\delta, \Delta}(x)$, for all $x
\in \{0,1\}^*$ and $\Delta > 0$. We set $\Delta_n=\epsilon n$. Then
the conditions of Theorem \ref{main-theorem} are satisfied and applying
(\ref{main-ineq}) we arrive at (\ref{implication-of-Thm10}). $\Box$

\section{Conclusions}\label{Sec:Conclusion}
In this contribution we studied the notion of plain effective
complexity, which is assigned to a given string, within the context of
an underlying stochastic process as model of the string generating
mechanism. In \cite{ECieee} we have shown that strings which are
called ``non-stochastic'' in the context of Kolmogorov minimal
sufficient statistics have large value of plain effective complexity.
The existence of such strings has been proven by G\'acs, Tromp and
Vit\'anyi in \cite{GacsTrompVitanyi}. Here, our aim was to understand
how properties of the stochastic process such as ergodicity and
stationarity influence the effective complexity of corresponding
typical realizations. Is it possible that the prefixes of a typical
process realization represent a sequence of finite strings in
increasing lenght $n$ that eventually have a high or moderate value of
effective complexity? Our main theorem refers to stationary and in
general {\it non computable} processes. It proves that modelling the
regularities of strings by computable ensembles with total information
that is allowed to excess the string's Kolmogorov complexity up to a
linearly growing amount $\epsilon n$ with an {\it arbitrary small}
$\epsilon >0$ is sufficient for typically generating non-complex
strings.

The value $\epsilon n$ plays the role of a parameter in the concept of
effective complexity. In order to have a notion that is independent
of this parameter we introduced coarse effective complexity. It
corresponds to coarse sophistication introduced by Antunes and Fortnow
in \cite{soph} and modifies effective complexity by incorporating the
parameter as a further minimization argument. Our result on effective
complexity has a direct implication on the asymptotic behaviour of
coarse effective complexity of typical realizations of a stationary
process. The main statement in this context demonstrates the utility
of the linear parameter scaling which we have considered. Moreover, it
allows to analyse the interplay between the complexity of a stochastic
process and the complexity of its typical realizations. In particular,
it demonstrates that, in order to have a notion of effective
complexity that also reflects the complexity of a stochastic process,
further modifications of plain effective complexity are necessary, for
instance introduction of appropriate constraints. This possibility is
in line with Gell-Mann and Lloyd's suggestion in
\cite{EffectiveComplexity} which we discussed in the Introduction.

Finally, we point out that continuing our previous work \cite{ECieee}
we have formulated our results for the concept of effective complexity
only. However, in line with the general equivalence statements obtained
in the literature, cf. Section V in \cite{Vitanyi} or Lemma 20 in
\cite{ECieee}, it should be possible to reformulate our main theorem in the more
general context of algorithmic statistics. Indeed, our upper bound on
effective complexity of typical process realizations is derived in
terms of computable ensembles that are uniform distributions on finite
sets (universally typical subsets). This demonstrates the close
relation in particular to the concept of Kolmogorov minimal sufficient
statistics which refers to the model class of finite sets.

\section*{Acknowledgements}

The authors would like to thank colleagues at the MPI MiS, in
particular Eckehard Olbricht, Wolfgang L\"ohr and Nils Bertschinger
for their interest and helpful discussions. This work has been
supported by the Santa Fe Institute.

\end{document}